\documentclass[12pt,a4paper]{article}
\usepackage{amsmath,amsfonts,amssymb,graphicx,epsf,lscape,latexsym}

\def\ds{\displaystyle}
\def\bea{\begin{array}{c}}
\def\ea{\end{array}}
\def\be{\begin{equation}\bea\ds}
\def\ee{\ea\end{equation}}

\def\d{{\rm d}}
\def\t{\tau}
\def\b{\beta}
\def\mb{\mu_\beta}
\def\Mu{{\cal{M}}}

\def\Sch{Schr\"odinger }
\def\RN{Reissner-Nordstr\"om }
\def\NMT{Null Melvin Twist }

\begin{document}

\begin{titlepage}

\begin{flushright}
{TAUP-2905/09}\\
{ITEP-TH-62/09}
\end{flushright}

\vspace{0.5cm}

\begin{center}
{\Large {\bf Stability of Asymptotically Schr\"odinger  RN Black Hole
\vskip 0.3cm
and Superconductivity}} \\

\vskip1.0cm {\large Stefano Cremonesi$^{a}$, Dmitry Melnikov$^{a,b}$ and Yaron Oz$^{a}$}
\vskip1cm
{\itshape
$^{a}$Raymond and Beverly Sackler School of Physics and Astronomy,\\
Tel Aviv University, Ramat Aviv 69978, Israel\\
\vspace{0.5cm}
$^{b}$Institute for Theoretical and Experimental Physics,\\
B.~Cheremushkinskaya 25, 117259 Moscow, Russia\\
}

\end{center}

\vspace{2cm}
\begin{abstract}
We perform a perturbative (near-critical) analysis of the stability of an asymptotically  \Sch \RN black hole with respect to generation of charged scalar hair. We find that apart from the expected instability at low temperatures typical of holographic models of superconductivity, in the presence of certain operators a similar instability appears as well at high temperatures. We propose that the reason for such a phase diagram could be due to peculiar features of the dual gauge theory or the failure of the model to provide a consistent holographic dual of a non-relativistic superconductor.
\end{abstract}

\end{titlepage}


\section{Introduction}

Asymptotically Schr\"odinger spacetimes with four noncompact dimensions \cite{Son:2008ye,Balasubramanian:2008dm} can be embedded in string theory \cite{Herzog:2008wg,Maldacena:2008wh,Adams:2008wt} by applying to asymptotically $AdS_5\times X_5$ solutions in type IIB string theory a solution-generating technique known as the Null Melvin Twist \cite{Bergman:2001rw,Alishahiha:2003ru,Gimon:2003xk,Hubeny:2005qu,Hubeny:2005pz}.

To be applied, the procedure requires two isometries, parameterized by $y$ and $\phi$, and involves a sequence of operations: a boost of rapidity $\gamma$ along $y$, T-duality along $y$, twist (shift) $d\phi\to d\phi+\alpha dy$, T-duality along $y$, boost of rapidity $-\gamma$ along $y$; finally, one has to take a double scaling limit $\gamma\to\infty$, $\alpha\to 0$, keeping $\beta=\frac{1}{2}\alpha e^\gamma$ fixed.

In \cite{Herzog:2008wg,Maldacena:2008wh,Adams:2008wt} the $y$ direction is one of the spacelike Minkowski directions inside $AdS_5$, and the $\phi$ direction can be chosen to be the Reeb vector direction in a Sasaki-Einstein manifold $X_5$.
The effect on the dual $3+1$ dimensional supersymmetric relativistic quantum field theories is to introduce a nonlocal dipole deformation along a null field theory direction, which is then compactified, leading to a twisted DLCQ of the relativistic field theory. The twist involves the R-symmetry of the parent $\mathcal{N}=1$ SCFT dual to the $AdS_5\times X_5$ background of type IIB string theory.
The resulting field theory on the remaining $2+1$ noncompact spacetime dimensions has Schr\"odinger symmetries, and the momentum along the null circle is dual to the particle number in the nonrelativistic field theory.

The Null Melvin Twist transforms a black hole solution with $AdS_5$ asymptotics into a black hole solution with Schr\"odinger asymptotics. This procedure was applied to $AdS_5$-Schwarzschild black holes that arise from non-extremal D3 brane backgrounds in \cite{Herzog:2008wg,Maldacena:2008wh,Adams:2008wt}. More recently, Reissner-Nordstr\"om (RN) and Kerr-Newman black holes in asymptotically Schr\"odinger spacetimes were generated \cite{Adams,Imeroni:2009cs} by applying the Null Melvin Twist to RN-$AdS_5$ black holes arising from non-extremal backgrounds of rotating D3 branes \cite{Chamblin:1999tk,Cvetic:1999xp}.

In the context of holographic models of superconductivity%
\footnote{Strictly speaking, in the field theory a global $U(1)$ is spontaneously broken, thus describing a charged superfluid. However, as in Landau-Ginzburg theory, this is enough to describe superconductivity, which occurs after the global $U(1)$ is weakly gauged.}
 \cite{Hartnoll:2008vx,Hartnoll:2008kx}, a RN black hole is the dual description of the normal (high temperature) phase of a superconductor; as the the background is cooled down, at some critical temperature a phase transition occurs to a black hole solution with charged hair, which describes the superfluid phase where a $U(1)$ symmetry is spontaneously broken by the VEV of a charged operator.%
\footnote{In this paper we will study scalar hair, which corresponds to $s$-wave superfluidity.}

Lately, holographic models of $(3+1)$- and $(2+1)$-dimensional relativistic superconductors have been embedded in type IIB string theory and M theory \cite{Gubser-Herzog-etal,Gauntlett:2009dn}. One could think of applying the Null Melvin Twist machinery to the type IIB models of holographic superconductors with $AdS_5$ asymptotics. However, in such a construction the operator that acquires a VEV in the $(3+1)$-dimensional $\mathcal{N}=1$ SCFT is the lowest component of a chiral superfield which is a linear combination of the superpotential and the glueball superfield, which has R-charge $2$. Correspondingly, the background in the superfluid phase breaks the isometry of the Reeb vector. As a consequence, one cannot apply to that solution the usual Null Melvin Twist that uses the universal Reeb vector direction as the second isometry.

If there is another $U(1)$ isometry in the Sasaki-Einstein manifold $X_5$, one can still apply the Null Melvin Twist to the type IIB holographic superconductor of \cite{Gubser-Herzog-etal}, but the result will depend on details of $X_5$.

Instead of tackling the problem top-down, an alternative option is that of studying the issue at the level of the 5d effective action written down in \cite{Adams:2008wt}, which is a consistent truncation of type IIB string theory and possesses a RN-Schr\"odinger black hole solution. One can then add by hand a charged scalar to this action and look for solutions of the equations of motion: we will add a minimally coupled scalar with a mass term. This is the minimal approximation we can imagine to a would-be effective action arising from a consistent truncation of type IIB string theory.
The 5d action of \cite{Adams} is quite cumbersome, and finding solutions to the full system of equations of motion of this action plus a charged massive scalar looks prohibitive, therefore we resort to the simplest possible approximation of studying the massive charged scalar as a probe in the charged black hole background of \cite{Adams}. In other words, similarly to what was done in \cite{Gubser} for $AdS_4$, we study the issue of the onset of the charged superfluid instability: we would like to determine for which values of the thermodynamical parameters marginally stable modes of the charged scalar appear. We will scan the parameter space of the model, varying the mass and charge of the scalar as well.

The paper is organized as follows. In section~\ref{setup sec} we briefly discuss the background solution and its thermodynamics. We start section~\ref{prelim sec} from establishing scaling symmetries of the system and continue with a discussion of the boundary conditions. An important observation there is that for fixed dimension of the dual operator the mass of the bulk field is a function of thermodynamical parameters. In section~\ref{numerics} we solve numerically the probe field equation searching for marginally stable modes. We find the critical lines of instability for a given set of operators in a scaling invariant phase space and analyze them in several cases. In section~\ref{numerics1} the temperature is varied at fixed chemical potentials. For operators with $\Delta\geq4$ we find a low temperature instability if a ratio of chemical potentials is not too small. For operators with $\Delta<4$ besides the low temperature one there is also a high temperature instability.
 For large ratio of chemical potentials low and high $T$ critical points merge and the background becomes unstable for all temperatures. In section~\ref{numerics2} we analyze the origin of the high temperature instability by exploiting the reformulation of the differential equation for the probe field as a zero energy \Sch problem. In section~\ref{sec scalings} we consider other scalings of thermodynamical variables and find that more generally a high $T$ instability is observed if $m^2L^2|_{T\to\infty}<0$. We summarize and discuss our findings in section~{\ref{sec conclusions}}.


\section{Setup}
\label{setup sec}
The five-dimensional asymptotically \Sch charged black hole background described in~\cite{Adams} contains the metric and a U(1) gauge field as well as a massive vector field and the dilaton. In the minimal scenario, which we motivated above, the probe scalar field is only coupled to the metric and the massless vector. The background form of the metric is given by
\be
\label{background metric}
\d s^2 = \frac{L^2K^{1/3}}{z^2}\left[- \frac{f\,\d \t^2}{K} - \frac{\b^2 f L^4}{z^2 K}\left(\d\t+\d y\right)^2 + \frac{\d y^2}{K}+ \frac{\d z^2}{f}+\d {\vec{x}}^{\,2} \right]\;,
\ee
while the vector field is simply
\be
\label{background potential}
A = \Phi\,\d\t\;, \qquad \Phi = \mu\left[1-\left(\frac{z}{z_h}\right)^2\right]\;.
\ee
Here $\b$ is the parameter of the Null Melvin Twist as in the above discussion. It is related to the particle number in the dual field theory, while $\mu$ is the chemical potential of the $U(1)$. The metric is written in the Poincar\'e patch with asymptotic boundary at $z=0$. The following functions and parameter have been introduced:
\be\label{functions}
\begin{split}
f &= 1 + (1+Q^2)\left(\frac{z}{z_h}\right)^4-Q^2\left(\frac{z}{z_h}\right)^6 \;, \\
K &= 1 + \frac{\b^2 L^4z^2}{z_h^4}\left[1+Q^2\,\frac{\Phi}{\mu}\right] \;,\\
Q &= \frac{2}{3}\,\mu z_h\;.
\end{split}
\ee
In these coordinates the horizon is at $z=z_h$ and the background before the \NMT coincides with the one of RN black hole in the conventions of~\cite{Gubser-Herzog-etal}.

In the above parametrization the relevant thermodynamical parameters take the following form. The temperature of the black hole is
\be
\label{temperature}
T=\frac{1}{\b L}\,\frac{\left|\,2-Q^2\right|}{2\pi z_h}\;,
\ee
so the black hole becomes extremal for $Q^2=2$. Since $0\leq z\leq z_h$ we should constrain $0\leq Q\leq\sqrt{2}$. The entropy density $s$ and the chemical potential $\mu_\b$ for the particle number are defined respectively by
\be\label{entropydens_chemicalparticlenumber}
4G_5\,s=\frac{L^3}{z_h^3}\;,\qquad \mu_\b = -\frac{1}{2\b^2L^3}\;.
\ee
With these relations at hand one can instead express the background solution in terms of the thermodynamical parameters $\mu$, $s$, $T$ and $\mu_\beta$, although the resulting expressions are somewhat involved.

To see if the dual gauge theory exhibits superconductivity (more properly, the existence of a charged superfluid phase) we have to investigate the stability of such a background with respect to generation of charged (scalar) hair. We will closely follow a similar analysis of the relativistic case performed by Gubser~\cite{Gubser}. In what follows we will be looking for marginally stable (zero) modes of the Klein-Gordon equation of the probe scalar field in the background (\ref{background metric}-\ref{background potential}), that is solutions of the equation
\be
\label{Klein-Gordon}
\frac{1}{\sqrt{-g}}\,\partial_z\left(\sqrt{-g}g^{zz}\partial_z\psi\right) - (m^2 + q^2 g^{\t\t}\Phi^2)\psi =0\;.
\ee
The explicit form of this equation can be found in the appendix~(\ref{Klein-Gordon2}). In the above equation the probe field is characterized by its mass $m$ and charge $q$, which are related to the dimension and charge of the dual operator. We will keep the latter fixed as parameters of the dual gauge theory.

An instability of the background~(\ref{background metric}-\ref{background potential}) related to superconductivity arises if at given values of the thermodynamical parameters there exists another background with the same asymptotics but lower free energy and non-trivial profile of scalar field. The asymptotic value of this scalar field at the boundary corresponds to the condensate of the dual operator. Thus the new background breaks the $U(1)$ symmetry and the dual theory becomes superconducting. We will show that in a certain range of the thermodynamical parameters the background is indeed unstable. However the approximation we are using does not allow us to compute the condensate itself. Neither it is possible to prove that the phase transition is second order. That analysis would require to solve the full coupled system of equations of motions.

The plan is to search for normalizable modes of~(\ref{Klein-Gordon}) in the background~(\ref{background metric}-\ref{background potential}) while varying certain scaling invariant parameters. Depending on the way one fixes thermodynamical parameters and on the operator content of the theory one can find an instability at low temperature, high temperature or both. This will be demonstrated by a numerical analysis in section~\ref{numerics}, but before we will discuss scaling symmetry and some analytical facts about the background.


\section{Preliminary Analysis}
\label{prelim sec}

\subsection{Scaling Invariance}
\label{scaling sec}
The asymptotically \Sch charged black hole background is associated with two scaling symmetries, which are the symmetries of the truncated five dimensional action of~\cite{Adams}. These can be described as specific rescalings of all variables and parameters $p_i\to \lambda^{\alpha_i} p_i$. The two scaling symmetries are parameterized by independent transformations $z\to \lambda^{\alpha_z}z$ and $L\to\lambda^{\alpha_L}L$, accompanied by
\be
\label{scalings}
\alpha_\mu = -\alpha_z\;, \qquad \alpha_\b = \alpha_z-2\alpha_L\;,\qquad \alpha_m = -\alpha_L\;, \qquad \alpha_q=0\;, \qquad \alpha_{G_5}=3\alpha_L\;.
\ee
Therefore the problem only depends on scaling invariant quantities.

Scaling symmetries are related to the redundancy of the description of the system with parameters $\mu$, $\b$, $L$, and $z_h$ or equivalently in terms of $\mu$, $\mb$, $T$ and $s$. A convenient way to fix the symmetries would be to rescale $\mu\to 1$ and $\b T\to 1$. It can be seen that after this rescaling the new quantities can be expressed just in terms of two scaling independent and dimensionless parameters
\be
\label{P,Q}
Q=\frac{2}{3}\,\mu z_h, \qquad {\rm and} \qquad P=\frac23\,\b\mu L^2\;.
\ee
In particular the equation~(\ref{Klein-Gordon}) written in terms of $P$ and $Q$ and the scaling invariant coordinate $w=z/z_h$ takes the form~ \eqref{Klein-Gordon2}-\eqref{A_B} presented in the appendix. This is also easily seen after (\ref{Klein-Gordon}) is recast in the form of a zero energy \Sch problem, as in \eqref{Schroed_problem}.
The \Sch potential is expressed in terms of $P$ and $Q$ in~(\ref{potential}).

Note also that the mass of the probe field is not an invariant parameter and therefore the equations will also depend on the combination $m^2L^2$ and the scaling invariant quantity $q$.


\subsection{Boundary conditions}
\label{bc section}

Let us first discuss the asymptotic behavior of solutions to~(\ref{Klein-Gordon}). Here we will use a scaling invariant radial coordinate $w=z/z_h$ and express the parameters in terms of $P$ and $Q$.  At the boundary ($w\to 0$), the equation takes the asymptotic form
\be
\label{KG asymptotic b}
\psi'' - \frac{3}{w}\,\psi' - \frac{1}{w^2}\left(m^2L^2+ \frac{9}{4}\,q^2P^2\right)\,\psi =0\;.
\ee
It has the general solution
\be
\psi_{(1)}w^{2+\alpha}+ \psi_{(2)}w^{2-\alpha},\qquad \alpha = \sqrt{4+m^2L^2+\frac94\,q^2P^2}\;.
\ee
Typically normalizable solutions are the ones with subleading behavior at the boundary: $\psi\sim w^{2+\alpha}$. However if $0<\alpha<1$, the solution $\psi\sim w^{2-\alpha}$ is also normalizable. In that regime one has to choose between two possible quantizations. The dimensions of the dual operator in these two cases are $\Delta=2\pm\alpha$, respectively. We notice that contrary to the asymptotically $AdS$ case, if one keeps the dimension and the charge of the dual operator fixed then the bulk mass of the scalar particle becomes parameter dependent:
\be
\label{m^2L^2_1}
m^2L^2 = \Delta(\Delta-4)-\frac{9}{4}\,q^2P^2\;.
\ee
At the horizon $w\to 1$ the equation (\ref{Klein-Gordon}) reduces to
\be
\psi'' + \frac{\psi'}{w-1} - \frac12\,\frac{m^2L^2\left(1+P^2/Q^2\right)^{1/3}}{(Q^2-2)}\,\frac{\psi}{w-1}=0\;.
\ee
The regular solution at the horizon can be expanded in series
\be
\psi_{\rm reg} = 1 + \frac12\,\frac{m^2L^2\left(1+P^2/Q^2\right)^{1/3}}{(Q^2-2)}\,(w-1)+ {\rm O}\left[(w-1)^2\right].
\ee
Here we normalized the function to satisfy $\psi(1)=1$.

It is also useful to obtain the near horizon expansion in the extremal case $Q^2=2$. Then the equation reads
\be
\psi'' + \frac{2\,\psi'}{w-1} + \frac{1}{24}\left(3q^2-2 m^2L^2 \left(1+\frac12\,P^2\right)^{1/3}\right)\frac{\psi}{(w-1)^2}=0\;.
\ee
In the presence of a low temperature instability the solution should be infinitely oscillating near the horizon in the extremal limit. This happens if
\be
\label{zero Tc bound1}
\left(1+\frac32\,P^2\left(1+\frac12\,P^2\right)^{1/3}\right)q^2 > 2 + \frac23\,\Delta(\Delta-4)\left(1+\frac12\,P^2\right)^{1/3}\;.
\ee
Notice that for $\b=0$ ($P=0$) we recover the relativistic condition derived in~\cite{Gubser-Herzog-etal}. The above condition can also be written in terms of the thermodynamical parameters at zero temperature,
\be
\label{zero Tc bound2}
\begin{split}
 2 &- \left(1-\frac{(4G_5s)^{1/3}}{\sqrt{2}}\,\frac{\mu}{\mu_\b}\left(1 - \frac{(4G_5s)^{1/3}}{3\sqrt{2}}\,\frac{\mu}{\mu_\b}\right)^{1/3}\right)q^2  +\\
&+ \left.
\frac23\,\Delta(\Delta-4)\left(1- \frac{(4G_5s)^{1/3}}{3\sqrt{2}}\,\frac{\mu}{\mu_\b}\right)^{1/3}\right|_{T=0}<0\;.
\end{split}
\ee
This condition tells that if the dimension of the operator is too large compared to its $U(1)$ charge it cannot condense at low temperature.


\section{Numerical Analysis}
\label{numerics}

\subsection{Superconducting instability}
\label{numerics1}

In this section we would like to study the behavior of the Einstein-Maxwell-scalar system with fixed dimension $\Delta$ and charge $q$ of the dual operator as the temperature is lowered at fixed chemical potentials. Due to the scaling symmetries (\ref{scalings}) the phase diagram will be 2-dimensional, parameterized by a pair of scaling invariant coordinates. The parameters $P$ and $Q$ introduced above are convenient for the analysis of equations of motion, however they are not as conveniently related to the thermodynamical parameters. To parameterize the phase space we will rather use their combinations
\be
\label{P,Q, tilde}
\tau\equiv \frac{P}{2\pi Q}\left(2-Q^2\right)=\frac{T}{-2\mb}\;, \qquad {\rm and} \qquad \Mu \equiv\frac{3}{2}\,\frac{P^2}{Q}=\frac{\mu}{-2\mb}\,(4G_5 s)^{1/3}\;.
\ee

We will search for instability by solving numerically the equation~(\ref{Klein-Gordon}) paying attention to normalizable modes. Since we are interested just in the onset of instability (critical temperature) we will be looking for normalizable solutions with no nodes except the one at the boundary. For that we will vary the temperature $T$ keeping $\mu$, $\mb$, $s$ fixed, that is vary $\tau$ in~(\ref{P,Q, tilde}), keeping fixed $\Mu$. In terms of the original parameters we will set
\be
\label{scaling1}
z_h=L\;, \qquad \b^2=L^{-3/2}\;,
\ee
so that $\tau=T$ and $\Mu=\mu$, and vary $L\in [0,3/(\sqrt{2}\mu)]$ for a given value of $\mu$ and given $\Delta$ and $q$.

For operators with $\Delta\geq 4$ we observe a picture similar to the relativistic case, e.g.~\cite{Gubser}. For sufficiently low $T=T_c^I$ (more generally, the scaling invariant combination $\tau$) we begin to observe a normalizable solution with no nodes and, as we lower the temperature further, more normalizable modes with more nodes. This is true of course if the chemical potential $\mu$ (more generally scaling parameter $\Mu$)\footnote{In the current and the next section we will practically  make no distinction between temperature $T$ and chemical potential $\mu$ and the scaling invariant parameters $\tau$ and $\Mu$ respectively. In later sections we will rather use $\tau$ and $\Mu$ to discuss phases in a more general setup than the one specified by~(\ref{scaling1}).} satisfies the bound~(\ref{zero Tc bound2}). Following~\cite{Gubser} we associate such a behavior of the background~(\ref{background metric}-\ref{background potential}) for $T<T_c^I$ with a low temperature instability with respect to the generation of a non-trivial profile of scalar field. From the point of view of the dual field theory, this instability corresponds to condensation of the dual operator and spontaneous breaking of the $U(1)$ symmetry, leading to superconductivity.

\begin{figure}[htb]
\begin{minipage}[b]{0.5\linewidth}
\begin{center}

\includegraphics[width=8.cm]{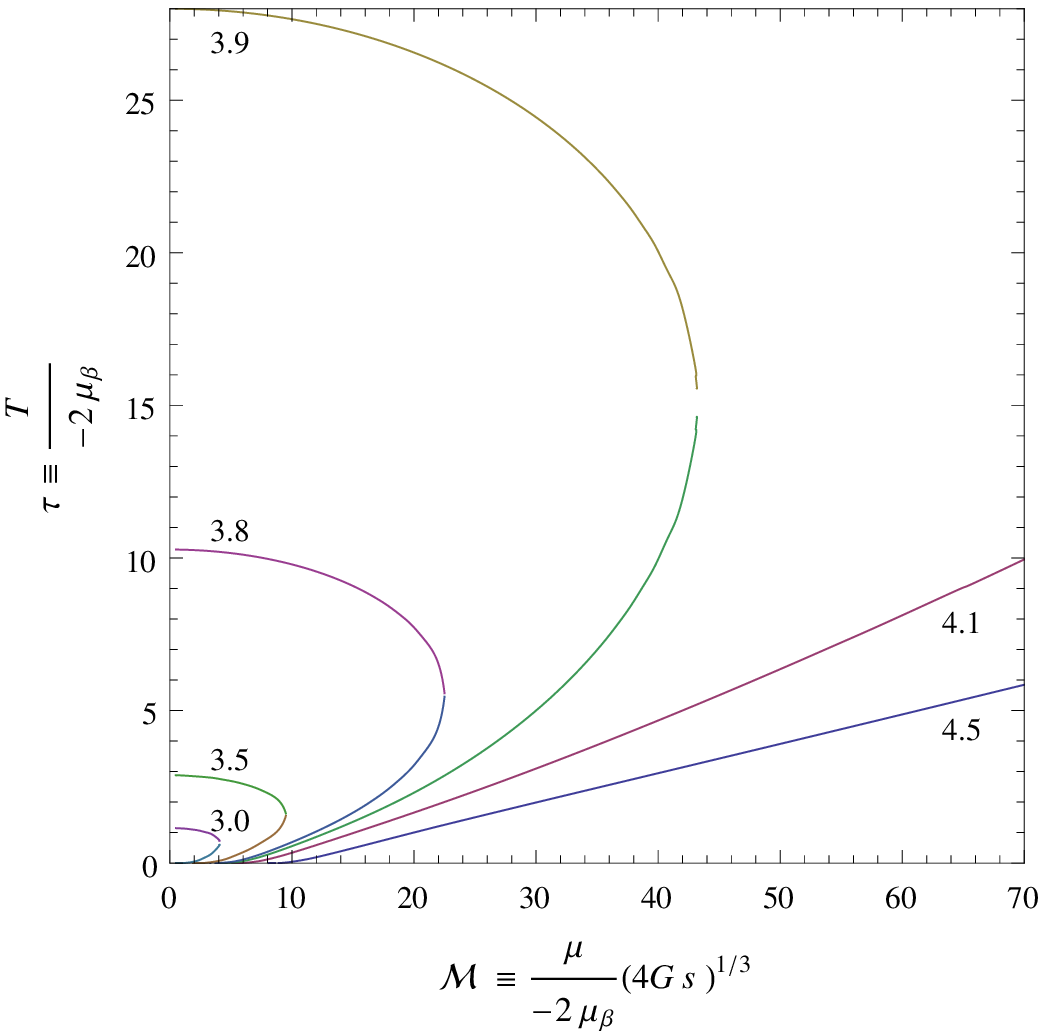}

(a)
\end{center}
\end{minipage}
\begin{minipage}[b]{0.5\linewidth}
\begin{center}

\includegraphics[width=7.9cm]{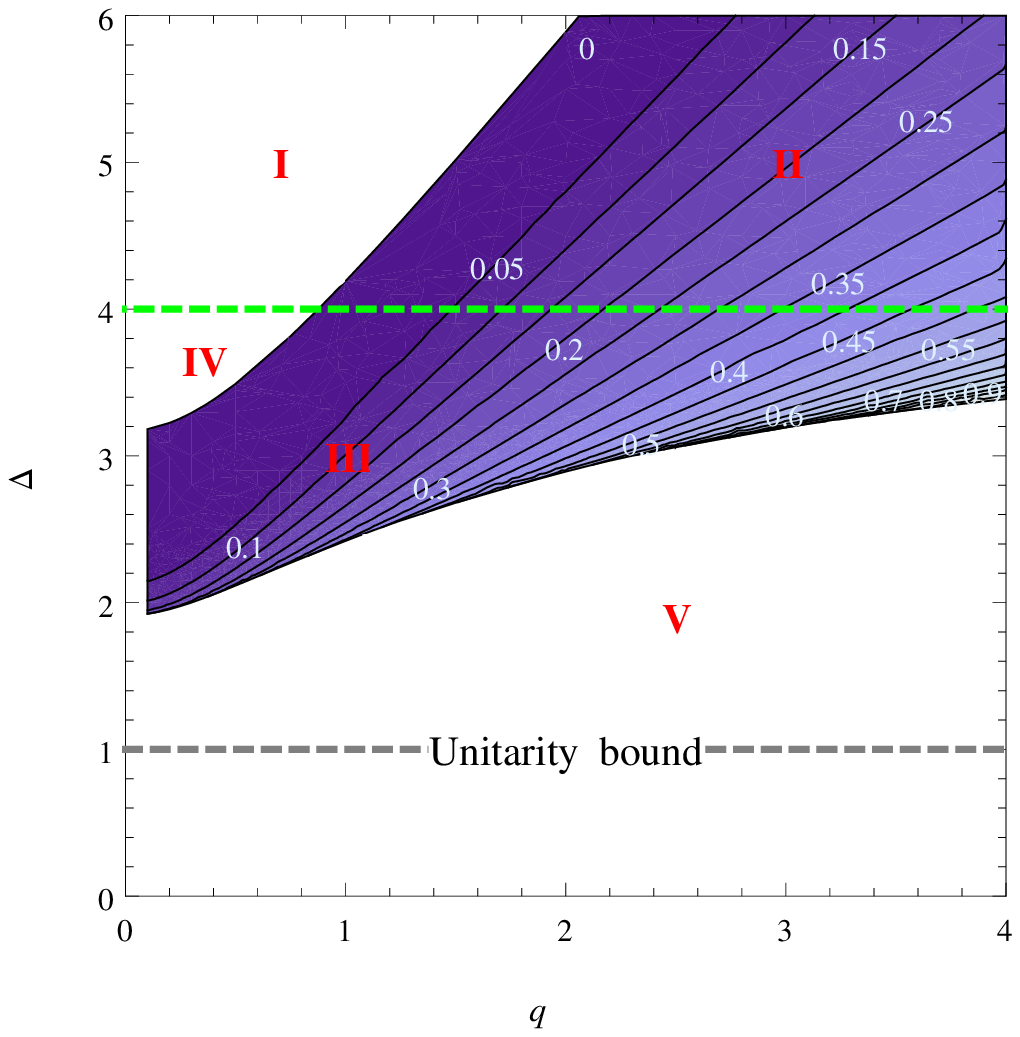}

(b)
\end{center}
\end{minipage}
\vspace{-0.6cm}
\caption{\small (a) Critical lines in the phase space for the low-temperature ($T_c^I$) and high-temperature ($T_c^{II}$) instability for operators with charge $q=1/2$ and different dimensions (shown on the plot). For $\Delta<4$, the derivative of $T_c(\mu)$ eventually diverges as one increases $\mu$, while for $\Delta>4$ it asymptotes a constant value. (b) A contour plot of the critical temperature $T_c^I$ as a function of the dimension and the charge of an operator for fixed chemical potentials ($4G_5s=1$, $\mu=1$, $\mb=-1/2$). The plot shows operators for which the background is: stable at any temperatures (region I), unstable at low temperature (II), unstable at both low and high temperatures (III), unstable only at high temperature (IV) or unstable at all temperatures (V). The shaded area is bounded from above by the zero $T_c$ bound (\ref{zero Tc bound2}) and from below by the overlap of high temperature and low temperature instabilities $T_c^I=T_c^{II}$.}
\label{mu-T phase space}
\end{figure}

In figure~\ref{mu-T phase space}(a), critical lines are shown for several operators with $\Delta>4$, as well as operators with $\Delta<4$. We will initially concentrate on the former, and consider the latter case in the next subsection. The lines start at zero $T_c$ for finite $\mu$ (or $\Mu$) in accordance with the bound~(\ref{zero Tc bound2}). For large $\mu$ the asymptotic form of the critical lines is almost linear. This fact was checked numerically for values of $\mu$ up to $\mu=10^3$ (for example see figure~\ref{D>4 asymptotics}).  In section~\ref{sec scalings} we will give an analytical derivation of the slope of the asymptotes.

In a given theory there may be several operators that condense at low temperature. However only the one with the highest critical temperature is relevant for the near-critical analysis of this work. It is thus useful to know the set of operators which can condense at least in principle. In asymptotically $AdS$ backgrounds such an investigation was made in~\cite{Gubser-Herzog-etal,Denef}. In those examples the region of condensing operators is finite and bounded by unitarity and BPS conditions as well as by zero $T_c$ bound similar to~(\ref{zero Tc bound2}).

In the asymptotically Schr\"odinger case, the critical temperature $T_c$ depends on one more parameter and for concreteness we will present the results only for a slice of the phase space with fixed $\mu$. Figure~\ref{mu-T phase space}(b) shows the critical temperature dependence on $\Delta$ and $q$ in such a case. Unlike relativistic examples, the region of condensing operators is not finite for chosen value of $\mu$, even if a BPS-like bound held for the non-supersymmetric background~(\ref{background metric}-\ref{background potential}). Therefore even large dimension operators can condense. Let us now discuss another important difference from the asymptotically $AdS$ backgrounds.


\subsection{High temperature instability}
\label{numerics2}

The situation is more peculiar in the case of the operators with dimension $\Delta<4$. As before, for sufficiently large $\mu$ the background is also unstable at low temperatures. However in this regime normalizable solutions with no nodes arise in pairs. This indicates that for $\Delta<4$ there are two critical temperatures $T_c^I<T_c^{II}$. Indeed we find normalizable solutions with multiple nodes both for $T<T_c^I$ and $T>T_c^{II}$. And while the number of nodes increases as the temperature is lowered in the first case, in the second case it increases when $T\to\infty$. Thus we have to conclude that in the presence of operators $\Delta<4$ the asymptotically \Sch black hole~(\ref{background metric}-\ref{background potential}) is unstable also at high temperatures.

If one increases the chemical potential $\mu$ the lower critical temperature $T_c^I$ will increase, while the higher critical temperature $T_c^{II}$ will decrease, eventually meeting each other at some value of $\mu=\mu_c$. For higher $\mu$ the low temperature and the high temperature unstable phases will overlap, thus rendering the background unstable for all temperatures. This feature of $\Delta<4$ operators is illustrated in figure~\ref{mu-T phase space}(a). The critical lines in this case bear the qualitative shape of semi-ellipses. In the $\Delta\to 4^-$ limit the critical temperature $T_c(\mu_c)$ is pushed up to eventually diverge for $\Delta=4$, and the semi-ellipses become monotonically increasing lines of small curvature.

It seems physically counterintuitive to relate the high temperature instability with the condensation of a charged scalar field and thus superconductivity. On the other hand we notice from figure~\ref{mu-T phase space}(a) that the high temperature and the low temperature unstable phases are smoothly connected, i.e. they are the same phase. We have no good field theory explanation for such an extension of the instability to high temperatures in $\Delta<4$ case, although we will present some considerations in the concluding section~\ref{sec conclusions}. Here we will analyze this effect from the point of view of the supergravity solution.

\begin{figure}[htb]
\begin{minipage}[b]{0.5\linewidth}
\begin{center}

\includegraphics[width=8.cm]{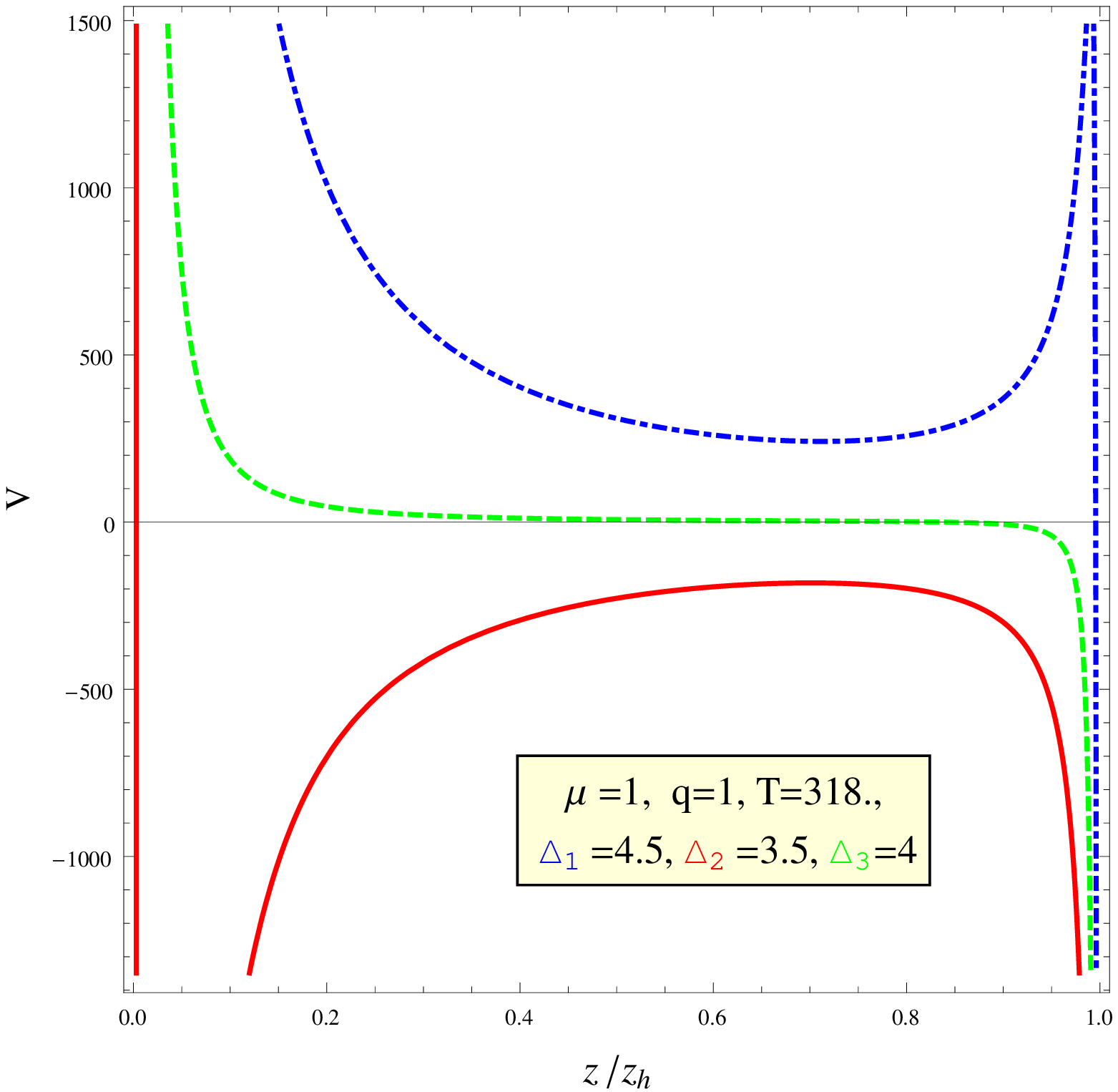}

(a)
\end{center}
\end{minipage}
\begin{minipage}[b]{0.5\linewidth}
\begin{center}

\includegraphics[width=8.cm]{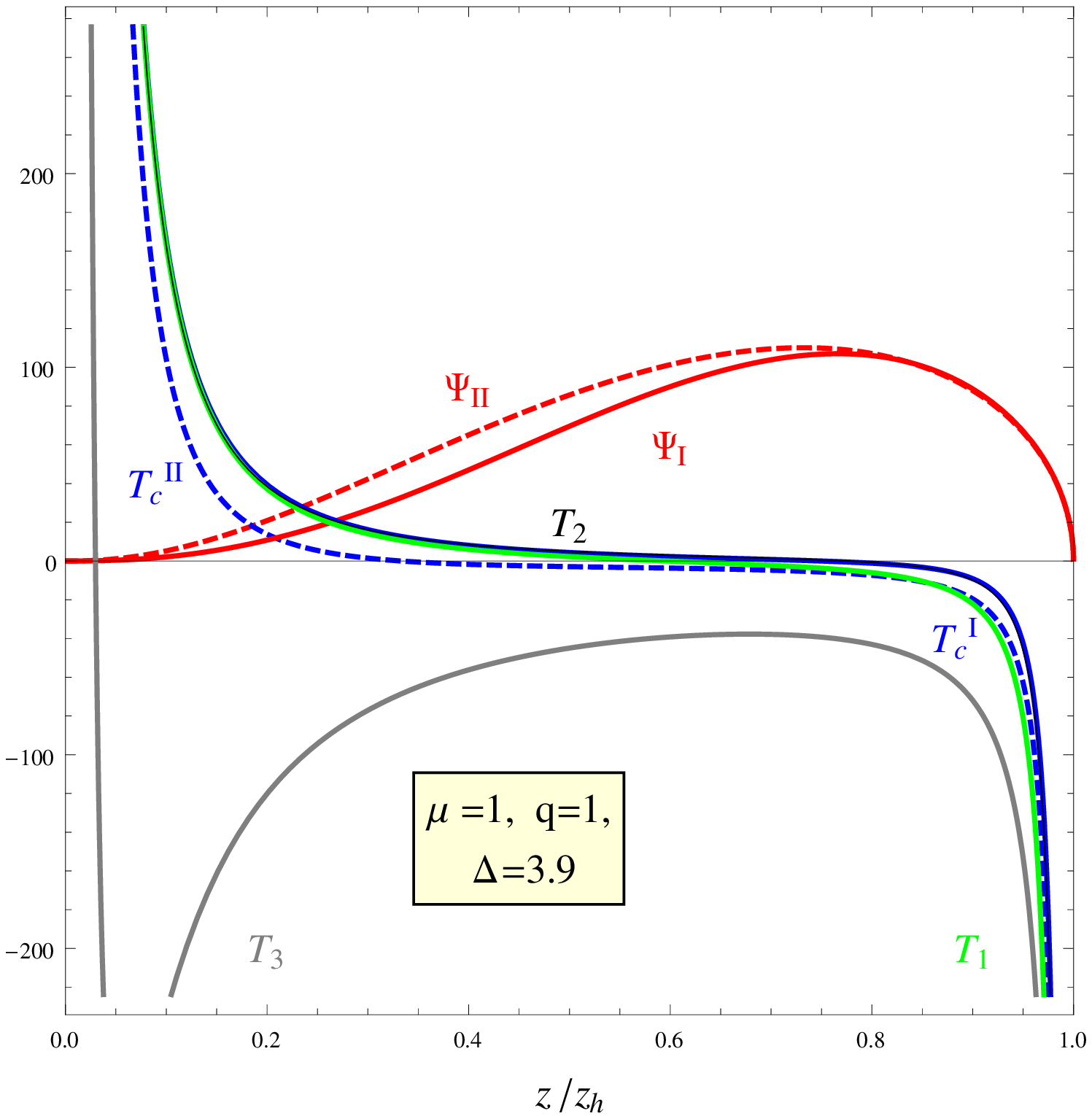}

(b)
\end{center}
\end{minipage}
\vspace{-0.6cm}
\caption{\small In the scaling $4G_5s=1$, $\mb=-1/2$: (a) high temperature form of potentials for $\Delta_1>4$ (dot-dashed), $\Delta_2<4$ (solid) and $\Delta=4$ (dashed). (b) Evolution of the potential with $T$ in $\Delta<4$ case. Here $T_1<T_c^I$, $T_c^I<T_2<T_c^{II}$ and $T_c^{II}<T_3$. $\Psi_I$ and $\Psi_{II}$ are normalizable modes that occur at $T_c^I$ and $T_c^{II}$ respectively.}
\label{potentials}
\end{figure}

One can understand the mathematical reason for the appearance of high temperature normalizable modes by looking at the temperature behavior of the potential in the Schr\"odinger problem defined in the appendix. Although the leading asymptotics of the potential at the horizon is the same for any $\Delta$, the near-horizon profile is still quite different for $T\to\infty$. In figure~\ref{potentials}(a) we plot sample potentials at high temperature for $\Delta<4$, $\Delta=4$ and $\Delta>4$. For $\Delta>4$ most of the potential (except for the very thin infinitely deep well close to the horizon) is above the $E=0$ level. It is reasonable to expect that zero energy normalizable solutions do not appear and we confirm that by numerical analysis. Conversely, for $\Delta<4$ most of the potential is below the $E=0$ level and there are normalizable modes in such a potential. $\Delta=4$ case is marginal. Most of the potential is again above the $E>0$ level, but there is no potential barrier, as in $\Delta>4$ case. Numerics does not reveal any normalizable solutions residing in the thin well close to the horizon.

The limits $z\to z_h$ and $T\to\infty$ do not commute, but one can still look at terms that give the most contribution in the $T\to\infty$ limit at small but finite distance from the horizon. The leading term in the potential~(\ref{potential}) in that limit is
\be
V= \frac{\pi^{2/3}\Delta(\Delta-4)}{2w^2(1-w^4)}\,T^{2/3}+ O\left[T^0\right]\;,
\ee
which is positive for $\Delta> 4$ and negative for $\Delta<4$, what explains the difference between the potentials in figure~\ref{potentials}(a). Thus whether high temperature normalizable modes appear or not depends on the sign of $\Delta(\Delta-4)$. We notice that in the high temperature limit of the scaling~(\ref{scaling1}) $\Delta(\Delta-4)=m^2L^2|_{T=\infty}$. (Recall that in the asymptotically \Sch case fixing $m$ and $\Delta$ is not the same thing.) Therefore there is a high temperature instability if $m^2L^2|_{T=\infty}<0$. We will argue later that the sign of $m^2L^2$ gives a stronger condition on operators, i.e. even $\Delta>4$ operators may lead to instability in certain limits,

In figure~\ref{potentials}(b) we display the evolution of the potential from low temperatures to high temperatures in a $\Delta<4$ example. The following observations can be made. In $T_c^I<T<T_c^{II}$ regime the potential is either too shallow or even mostly positive to allow for normalizable solutions. For low and high temperature normalizable solutions ($T_c^I$, $T_c^{II}$) the potentials (and the solutions) are quite similar although not precisely the same. We notice that for high $T$ there is a tendency for the particle to run away from the horizon.

The latter observation suggests that the high temperature instability is similar to the super-radiant instability of asymptotically flat black holes. In the parametrization used in~\cite{Gubser} an asymptotically flat black hole becomes super-radiant if $m^2<4q^2$, where as before $m$ and $q$ are the probe mass and charge. To compare its behavior to the present case we also plot the Schr\"odinger potential of equation~(9) in~\cite{Gubser}. The asymptotically flat case corresponds to taking the $L\to\infty$ limit. After setting $r_+=Q=1$, the Hawking temperature of the background is given by
\be
T=\frac{1}{4\pi}\,\left(k-\frac14\right), \qquad k\geq\frac14 \;.
\ee

\begin{figure}[htb]
\begin{minipage}[b]{0.5\linewidth}
\begin{center}

\includegraphics[width=7.8cm]{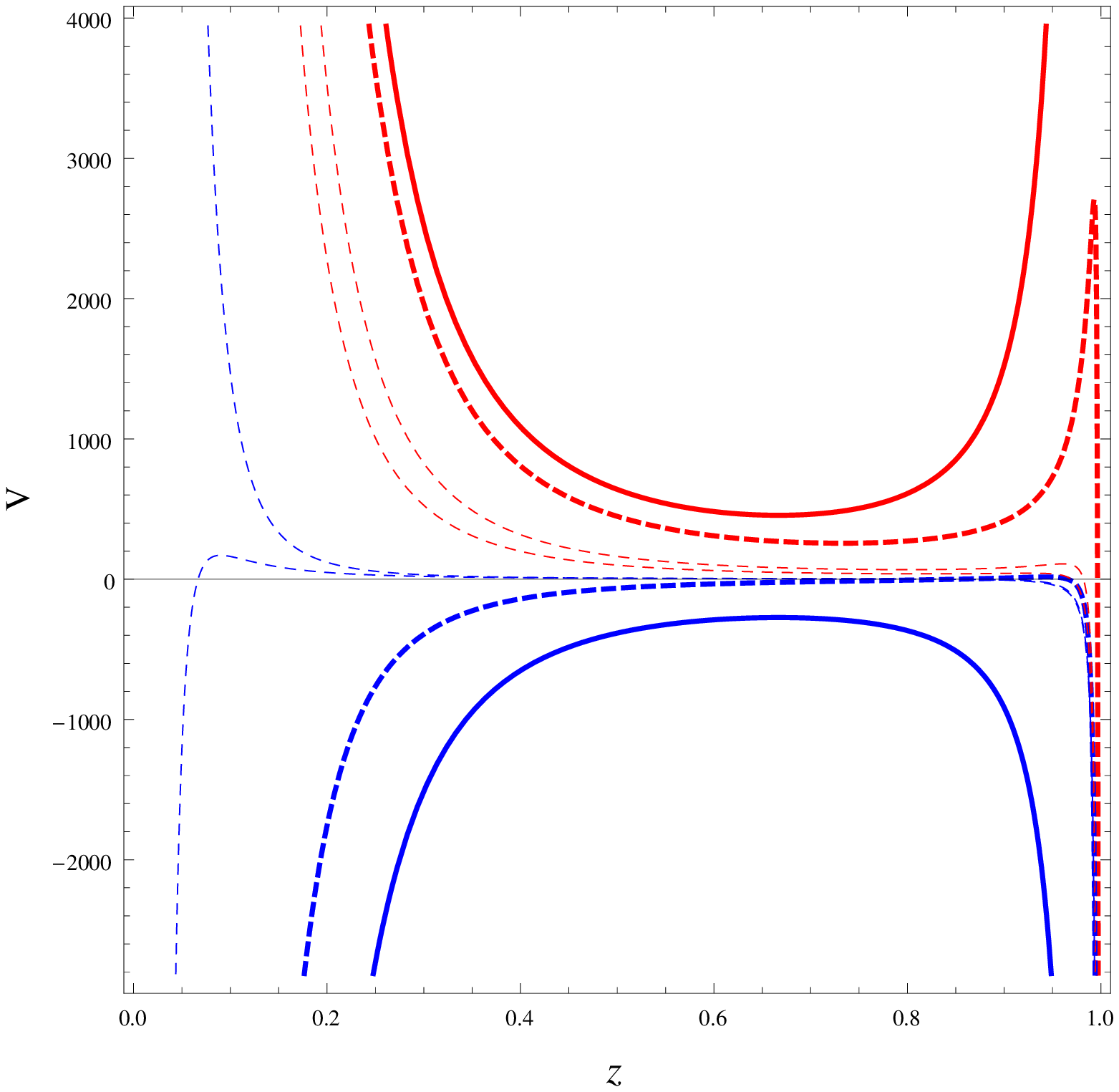}

(a)
\end{center}
\end{minipage}
\begin{minipage}[b]{0.5\linewidth}
\begin{center}

\includegraphics[width=8.cm]{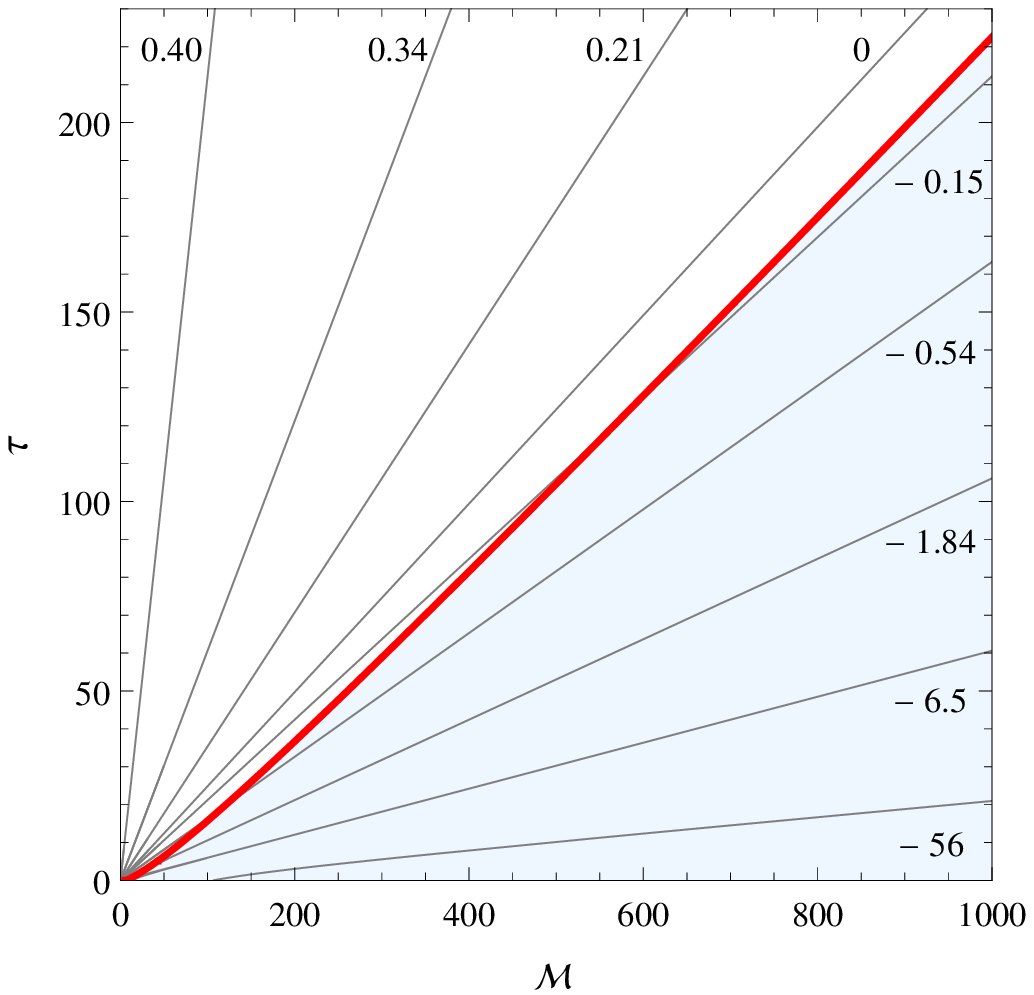}

(b)
\end{center}
\end{minipage}
\vspace{-0.6cm}
\caption{\small (a) Temperature evolution of the probe potentials in the background of normal (red) and super-radiant (blue) asymptotically flat black holes. Thick solid curves correspond to extremal black holes and dashed lines reflect the evolution with temperature raised. (b) A derivation of the asymptotic slope of the critical curve for operator $\Delta=4.1$, $q=0.5$ (thick red curve). Lines of fixed $P$ (gray) are parameterized by $Q$ and labeled by values of $m^2L^2$. For $m^2L^2<0$ lines end in the shaded unstable region. The slope of the critical curve asymptotes to the slope of $m^2L^2=0$ line.}
\label{D>4 asymptotics}
\end{figure}
In figure~\ref{D>4 asymptotics}(a) we plot the evolution of the potential with the temperature for both normal $m^2>4q^2$ and super-radiant asymptotically flat black holes (the original coordinate is rescaled to $z=r^{-1}$ for the sake of comparison). In the super-radiant case the potential develops an infinite well below a certain temperature. As the temperature decreases the well widens. When it is wide enough normalizable solutions start to populate it. Such normalizable solutions will be localized close to the boundary. In other words bulk particles will tend to escape from the black hole to the boundary, or the black hole will super-radiate.

One observes a similar tendency in the case of potential~(\ref{potential}) in figure~\ref{D>4 asymptotics}(a). Tachyonic probe particles will be driven away from the horizon if one raises the temperature. Although going to the large temperature limit is not justified within the near-critical approximation that we employ, this argument elucidates the mathematical reason for the instability at high temperatures from the bulk point of view. Super-radiance of the bulk solution at high temperatures is another factor that affects the stability.


\subsection{Interesting scalings}
\label{sec scalings}
In similar studies of asymptotically $AdS$ backgrounds in~\cite{Gubser-Herzog-etal,Gubser,Denef}, the phase space is one dimensional: the physics depends only on the ratio $T/\mu$. In the asymptotically Schr\"odinger case we have more freedom for possible experiments. For example we can choose such a scaling of parameters that the high temperature and low temperature instabilities of the previous section interchange. From the point of view of figure~\ref{mu-T phase space}(a) this would signify that we can fix $\Mu$~(\ref{P,Q, tilde}), but move in the direction of small $\tau$ while increasing $T$. For that it is sufficient to scale
\be
\mb\propto \mu \left(\frac{T}{\mu}\right)^{\alpha}\;, \qquad \alpha>1\;, \qquad \mu={\rm const}.\;,\qquad 4G_5\,s\propto \left(\frac{T}{\mu}\right)^{3\alpha}\;.
\ee
For this scaling of the parameters a critical point $\tau_c^{II}$ that previously corresponded to the onset of high temperature instability is at lower temperature than $\tau_c^I$. We notice that in such a regime the high temperature condensation is physically sensible, since one increases the chemical potentials faster then the temperature. The high $\tau$ instability is still mysterious, but as it now occurs at low temperatures we note that it looks even more similar to the super-radiant instability of the asymptotically flat black hole.

Another interesting example to consider is to look at the scalings of the invariant variables $Q$ and $P$. In particular consider a limit
\be
\label{P-const,Q->0}
P={\rm const.}\;, \qquad Q\to 0\;,
\ee
that is vary Q ($0<Q<\sqrt{2}$), keeping $P$ fixed. In terms of the phase space in figure~\ref{mu-T phase space}(a) this scaling corresponds to moving away from the origin along the curves shown in figure~\ref{D>4 asymptotics}(b). In terms of the thermodynamical variables one can understand this scaling as keeping $\mb$ fixed while increasing $\mu$ and $T$ at the same rate asymptotically.  The leading term of the Schr\"odinger potential~(\ref{potential}) in the limit~(\ref{P-const,Q->0}) is
\be
V = \frac{(P^2 w^2)^{1/3}\left[\Delta(\Delta-4)-9q^2P^2/4\right]}{2w^2(1-w^4)Q^{2/3}}+O(1)\;.
\ee
Thus we can expect normalizable modes in the small $Q$ limit if $\Delta(\Delta-4)-9q^2P^2/4$ is negative, i.e. when the probe particle is tachyonic, $m^2L^2<0$. Notice that it is always negative if $\Delta<4$.

Again the physical reason for the condensation is not intuitively clear in the limit~(\ref{P-const,Q->0}). The behavior of the potential is similar to the high temperature behavior in the previous section. Therefore we again interpret this as the effect of super-radiant behavior of the non-relativistic black hole.

In figure~\ref{D>4 asymptotics}(b) we label lines of fixed $P$ in the $(\tau,\Mu)$ phase space by the values of $m^2L^2$. In that figure we plot the critical curve for an operator with $\Delta>4$. As can be seen from the figure, the $Q\to 0$ limit is indeed unstable (shaded part of the phase space) if $m^2L^2<0$. Interestingly this observation allows us to find the slope of the critical curve. It is given by
\be
\left.\frac{2}{3\pi P}\right|_{m^2L^2=0}=\frac{q}{\pi\sqrt{\Delta(\Delta-4)}}\;.
\ee
The slope diverges if $\Delta=4$ which is consistent with our earlier observations about the $\Delta\to 4^-$ limit in figure~\ref{mu-T phase space}(a).


\section{Discussion and conclusions}
\label{sec conclusions}

In this paper we studied the question of stability of an asymptotically Schr\"odinger RN black hole against formation of charged scalar hair. The investigation was perturbative, that is we considered the response of the background to small fluctuations of a charged scalar field, ignoring backreaction on the background and self-interaction of the scalar field, which should be enough for establishing instability in a vast number of models including possibly string compactifications. Similarly to the situation in the asymptotically $AdS$ cases \cite{Hartnoll:2008vx,Hartnoll:2008kx,Gubser-Herzog-etal,Gauntlett:2009dn,Gubser,Denef}, the black hole is unstable at low temperature if the dual theory contains operators with $\Delta/q$ not too large (\ref{zero Tc bound2}). In the dual field theory this instability is attributed to condensation of the operator dual to the charged scalar field and consecutive breaking of the $U(1)$ symmetry, leading to superconductivity.

In the meantime the non-relativistic theory by construction has an additional thermodynamical parameter, the particle number chemical potential $\mb$. This makes the phase space of the theory and its physics richer.
The phase space of the asymptotically Schr\"odinger RN black hole~\cite{Adams} studied in this work is effectively two dimensional, i.e. it can be parameterized by two scaling invariant combinations of thermodynamical variables such as $\tau$ and $\Mu$ in~(\ref{P,Q, tilde}). A novel feature of the non-relativistic theory is a non-trivial pattern of stable and unstable phases in the phase space. In particular in an experiment with fixed chemical potentials $\mu$ and $\mb$ (fixed $\Mu$) one will typically observe a new (unstable) phase at low temperature for any operator satisfying~(\ref{zero Tc bound2}); however, one will also find a second instability at high temperatures if $\Delta<4$. It is easy to see this from figure~\ref{mu-T phase space}(a) where the critical lines of several operators are plotted.

The effect can be qualitatively understood from the behavior of the potential of a tachyonic probe field.\footnote{Here we call tachyonic the probe field satisfying $\Delta<4$ condition. This is consistent with a more general definition~(\ref{tachyon}) in the case when the temperature is varied at fixed chemical potentials.} At  high temperatures the potential develops a well close to the boundary, which becomes infinitely deep at infinite temperature (super-radiant-like behavior). As a result new normalizable modes, localized closer to the boundary, appear at high temperature.

This discussion can be generalized after noticing that the probe field potential also starts exhibiting a super-radiant-like behavior at large $\Mu$, in other words it becomes super-radiant at any point at infinity of the ($\tau$, $\Mu$) phase space.  Recall that in such a case also $\Delta>4$ operators could condense at large temperatures. A more general condition for the super-radiant-like instability is the existence of tachyonic particles in the sense
\be
\label{tachyon}
m^2L^2|_{\tau(\xi_\infty),\, \Mu(\xi_\infty)}\, < 0,
\ee
where $(\tau(\xi),\Mu(\xi))$ is a parametrization of the curve on the phase space, which corresponds to a choice of the scaling of the thermodynamical parameters.

The physical interpretation of the high $\tau$ instability is not entirely clear. On one hand the high $\tau$ unstable phase is smoothly connected to the small $\tau$ unstable phase along the critical line. On the other hand a condensation for growing temperature but fixed chemical potential is puzzling. We see several possibilities in resolution of this issue. Let us discuss them in more detail.

The first possibility is that the behavior of the critical lines in figure~\ref{mu-T phase space}(a), inferred from the gravity solution, is consistent with the dual field theory. The latter is a DLCQ of a dipole deformation of a relativistic 4-dimensional gauge theory, and our physical intuition may not work very well in that case. It is conceivable that  the observed behavior is related to the non-locality of the theory.

The second possibility is that the approximation that we are using fails to describe a consistent gravity dual in certain regions of the phase space. We see at least two ways this can happen. First of all, the Abelian Higgs model could provide a good description of superconductivity, but the action could take a more general form of non-linear sigma-model. This might change the phase diagram found in our approximation of a canonically normalized massive scalar. Secondly, one can entertain the possibility that the consistent gravity dual has other sorts of instability related to different fields than the charged scalars we have considered in our analysis. One would need to take into account effects of other fields coming from string compactifications, which also can generate new phases restoring the $U(1)$ symmetry at high temperatures or even drastically change the phase diagram.

Our conclusions could be tested by applying to the type IIB solution of~\cite{Gubser-Herzog-etal} a \NMT involving a $U(1)$ isometry (other than the R-symmetry) of the Sasaki-Einstein manifold that remains unbroken in such a solution.


\section*{Acknowledgments}

We are grateful to O.~Aharony, Y.~Dagan, C.~Herzog, D.~T.~Son, J.~Sonnenschein and A.~Yarom for useful discussions.
The work is supported in part by the Israeli
Science Foundation center of excellence, by the Deutsch-Israelische
Projektkooperation (DIP), by the US-Israel Binational Science
Foundation (BSF), and by the German-Israeli Foundation (GIF). The work of DM was also partly supported by the RFBR grant 07-02-01161 and the grant for Support of Scientific Schools NSh-3035.2008.2.

\appendix
\numberwithin{equation}{section}

\section{Equations and \Sch potential}
\label{appendix}
After plugging in equation \eqref{Klein-Gordon} the expressions for the components of the metric and vector field, the differential equation takes the form
\be
\label{Klein-Gordon2}
\psi''(w) + {\cal{A}}(w)\,\psi'(w) + {\cal{B}}(w)\,\psi(w) =0\;,
\ee
where we introduced the scaling invariant dimensionless coordinate $w=z/z_h$ and the functions
\be\label{A_B}
\begin{split}
{\cal{A}}(w)&=\frac{-3 - (1+Q^2)w^4 + 3Q^2w^6}{w(1-(1+Q^2)w^4+Q^2 w^6)}\\
{\cal{B}}(w)&= -m^2L^2\,\frac{\left[1+\frac{P^2}{Q^2}w^2(1+Q^2-Q^2w^2)\right]^{1/3}}{w^2(1-w^2)(1+w^2-Q^2w^4)} +\\
&+ \frac{9}{4}\,q^2\left(\frac{Q^2}{(1+w^2-Q^2w^4)^2}- \frac{P^2(1-w^2)}{w^2(1+w^2-Q^2w^4)}\right)\;.
\end{split}
\ee
$\cal{A}$ and $\cal{B}$ are written in terms of the two scaling invariant parameters $P$ and $Q$,
\be\label{P_Q_app}
Q=\frac{2}{3}\,\mu z_h\;, \qquad P=\frac23\,\b\mu L^2\;,
\ee
and the dimensionless squared mass $m^2L^2$, which is a function of $P$ and the dimension and charge of the operator:
\be\label{mL_app}
m^2L^2=\Delta(\Delta-4)-\frac{9}{4}\,q^2 P^2\;.
\ee
It turns useful to rewrite the equation~(\ref{Klein-Gordon2}) in the form of a static Schr\"odinger equation at zero energy. With the wave-function redefinition
\be\label{wave_funt_redef}
\psi(w)= e^{-\frac{1}{2}\int \d w {\cal{A}}(w)}\,\Psi(w) = \frac{w^{3/2}}{3\sqrt{\left(1-w^2\right)\left(1+w^2-Q^2w^2\right)}}\,\Psi(w)\;,
\ee
equation~(\ref{Klein-Gordon2}) is recast in the \Sch form
\be\label{Schroed_problem}
-\frac{1}{2}\Psi''(w)+V(w)\, \Psi(w) =0
\ee
with a potential
\be\label{potential}
\begin{split}
V(w) & \equiv \frac{1}{8}\left[2\mathcal{A}'(w)+\mathcal{A}(w)^2-4\mathcal{B}(w)\right]=\\
&=  m^2L^2\,\frac{(1+P^2w^2(1+Q^2-Q^2w^2)/Q^2)^{1/3}}{2w^2(1-w^2)(1+w^2-Q^2w^4)} +\\
&-\frac{9}{8}\,q^2\left(\frac{Q^2}{(1+w^2-Q^2w^4)^2}- \frac{P^2(1-w^2)}{w^2(1+w^2-Q^2w^4)}\right) +\\
&+ \frac{15-30 \left(1+Q^2\right) w^4+54 Q^2 w^6-\left(1+Q^2\right)^2 w^8-6 \left(Q^2+Q^4\right) w^{10}+3 Q^4 w^{12}}{8 w^2 \left(1-w^2\right)^2 \left(1+w^2-Q^2 w^4\right)^2}\;.
\end{split}
\ee



\begin{thebibliography}{0}

\bibitem{Son:2008ye}
  D.~T.~Son,
  Phys.\ Rev.\  D {\bf 78}, 046003 (2008)
  [arXiv:0804.3972 [hep-th]].

\bibitem{Balasubramanian:2008dm}
  K.~Balasubramanian and J.~McGreevy,
  Phys.\ Rev.\ Lett.\  {\bf 101}, 061601 (2008)
  [arXiv:0804.4053 [hep-th]].

\bibitem{Herzog:2008wg}
  C.~P.~Herzog, M.~Rangamani and S.~F.~Ross,
  JHEP {\bf 0811}, 080 (2008)
  [arXiv:0807.1099 [hep-th]].

\bibitem{Maldacena:2008wh}
  J.~Maldacena, D.~Martelli and Y.~Tachikawa,
  JHEP {\bf 0810}, 072 (2008)
  [arXiv:0807.1100 [hep-th]].

\bibitem{Adams:2008wt}
  A.~Adams, K.~Balasubramanian and J.~McGreevy,
  JHEP {\bf 0811}, 059 (2008)
  [arXiv:0807.1111 [hep-th]].


\bibitem{Bergman:2001rw}
  A.~Bergman, K.~Dasgupta, O.~J.~Ganor, J.~L.~Karczmarek and G.~Rajesh,
  Phys.\ Rev.\  D {\bf 65}, 066005 (2002)
  [arXiv:hep-th/0103090].

\bibitem{Alishahiha:2003ru}
  M.~Alishahiha and O.~J.~Ganor,
  JHEP {\bf 0303}, 006 (2003)
  [arXiv:hep-th/0301080].

\bibitem{Gimon:2003xk}
  E.~G.~Gimon, A.~Hashimoto, V.~E.~Hubeny, O.~Lunin and M.~Rangamani,
  JHEP {\bf 0308}, 035 (2003)
  [arXiv:hep-th/0306131].

\bibitem{Hubeny:2005qu}
  V.~E.~Hubeny, M.~Rangamani and S.~F.~Ross,
  JHEP {\bf 0507}, 037 (2005)
  [arXiv:hep-th/0504034].

\bibitem{Hubeny:2005pz}
  V.~E.~Hubeny, M.~Rangamani and S.~F.~Ross,
  Int.\ J.\ Mod.\ Phys.\  D {\bf 14}, 2227 (2005)
  [arXiv:gr-qc/0504013].

\bibitem{Adams} A.~Adams, C.~M.~Brown, O.~DeWolfe and C.~Rosen,
  arXiv:0907.1920 [hep-th].

\bibitem{Imeroni:2009cs}
  E.~Imeroni and A.~Sinha,
  JHEP {\bf 0909}, 096 (2009)
  [arXiv:0907.1892 [hep-th]].

\bibitem{Chamblin:1999tk}
  A.~Chamblin, R.~Emparan, C.~V.~Johnson and R.~C.~Myers,
  Phys.\ Rev.\  D {\bf 60} (1999) 064018
  [arXiv:hep-th/9902170].

\bibitem{Cvetic:1999xp}
  M.~Cvetic {\it et al.},
  Nucl.\ Phys.\  B {\bf 558} (1999) 96
  [arXiv:hep-th/9903214].


\bibitem{Hartnoll:2008vx}
  S.~A.~Hartnoll, C.~P.~Herzog and G.~T.~Horowitz,
  Phys.\ Rev.\ Lett.\  {\bf 101}, 031601 (2008)
  [arXiv:0803.3295 [hep-th]].

\bibitem{Hartnoll:2008kx}
  S.~A.~Hartnoll, C.~P.~Herzog and G.~T.~Horowitz,
  JHEP {\bf 0812}, 015 (2008)
  [arXiv:0810.1563 [hep-th]].

\bibitem{Gubser-Herzog-etal}
  S.~S.~Gubser, C.~P.~Herzog, S.~S.~Pufu and T.~Tesileanu,
  arXiv:0907.3510 [hep-th].

\bibitem{Gauntlett:2009dn}
  J.~P.~Gauntlett, J.~Sonner and T.~Wiseman,
  arXiv:0907.3796 [hep-th].

\bibitem{Gubser} S.~S.~Gubser,
  Phys.\ Rev.\  D {\bf 78}, 065034 (2008)
  [arXiv:0801.2977 [hep-th]].

\bibitem{Denef} F.~Denef and S.~A.~Hartnoll,
  Phys.\ Rev.\  D {\bf 79}, 126008 (2009)
  [arXiv:0901.1160 [hep-th]].


\end{thebibliography}
\end{document}